\documentclass[12pt]{article}

\usepackage{bbm,epsfig}

\newcommand{\be}{\begin{equation}}
\newcommand{\ee}{\end{equation}}
\newcommand{\ba}{\begin{eqnarray}}
\newcommand{\ea}{\end{eqnarray}}
\newcommand{\no}{\nonumber \\}

\newcommand{\LL}{{\cal L}}
\newcommand{\mnu}{{\cal M}_\nu}

\begin{document}

\title{
\normalsize \hfill CFTP/11-017 \\[1cm]
\LARGE On a possible relationship between lepton mixing \\
\LARGE and the stability of dark matter
}
\author{
L.\ Lavoura\thanks{E-mail: balio@cftp.ist.utl.pt} \\
\small Technical University of Lisbon,
Centre of Theoretical Particle Physics \\
\small Instituto Superior T\'ecnico, 1049-001 Lisboa, Portugal
}


\maketitle

\begin{abstract}
I comment on the proposal that
the stability of dark matter may be due to
an unbroken $\mathbbm{Z}_2$ symmetry
contained in the partially broken lepton flavour symmetry group.
I remark that
(1) there is no $\mathbbm{Z}_2$ symmetry apparent
in the lepton mass spectrum and in lepton mixing,
(2) predictive models of this type
may be constructed by using a lepton flavour symmetry group
with three inequivalent singlets,
to which the three left-handed-lepton gauge-SU(2) doublets are assigned,
and (3) some predictions for the lepton masses and mixings
are likely to be altered
by radiative contributions to the neutrino mass matrix. 
I construct two models of this type
in which the conserved $\mathbbm{Z}_2$
originates in a lepton flavour symmetry group $D_4$.

\vspace*{10mm}

\end{abstract}

\newpage

\section{Introduction}

A large proportion of the matter in the Universe is ``dark''
and must be composed of some non-baryonic,
neutral,
massive,
and stable---or with lifetime much larger than
the age of the Universe---particle or particles.
From the point of view of high-energy physics,
the stability,
on the cosmological time scale,
of dark matter suggests the existence of some {\em exact} symmetry,
most likely a $\mathbbm{Z}_2$ under which
dark matter is odd while normal matter is even.
If this is the case,
then one may assume this $\mathbbm{Z}_2$ \textit{ad hoc}\/ but we would like,
instead,
to find a compelling reason for its existence.

It has been suggested~\cite{valle} that the stability of dark matter
may be due to an unbroken $\mathbbm{Z}_2$ subgroup of a
spontaneously (partially) broken flavour symmetry group in the lepton sector.
In two specific models embodying this idea~\cite{valle,morisi},
an $A_4$ lepton flavour symmetry group
has been used.\footnote{The group $A_4$
had been used before
as a means to derive the apparent tri-bimaximal form of lepton mixing.
In the presently considered models,
though,
$A_4$ is used with other assignments of the matter fields
to its representations,
and does {\em not} lead to tri-bimaximal lepton mixing.}
The two models claim specific predictions
for the neutrino masses and mixings:
the first model~\cite{valle}
claims the predictions $m_3 = 0$
and $U_{e3} = 0$,\footnote{These predictions are common to
other models~\cite{m3Ue3} which do not use $A_4$.}
where $U$ is the lepton mixing matrix
and $m_3$ is the mass of the third light neutrino;
the second model~\cite{morisi}
apparently predicts $\left( \mnu^{-1} \right)_{\mu\mu}
= \left( \mnu^{-1} \right)_{\tau\tau}
= 0$,\footnote{These predictions had already been shown~\cite{I}
to be compatible with the experimental data on neutrino masses and mixings.}
where $\mnu$ is the effective light-neutrino Majorana mass matrix
in the basis where the charged-lepton mass matrix
$M_\ell = {\rm diag} \left( m_e, m_\mu, m_\tau \right)$
is diagonal.

The fact that two distinct models have been produced
with identical $A_4$ lepton flavour symmetry group
and identical $\mathbbm{Z}_2$ unbroken subgroup,
but with distinct predictions for lepton mixing,
makes one suspicious that the unbroken $\mathbbm{Z}_2$
may have little to do with the claimed predictions.
This suspicion marks the starting point of this investigation.

This paper is organized as follows.
In section~2 I show that there is no exact $\mathbbm{Z}_2$ symmetry
in the lepton masses and mixings;
therefore,
the exact $\mathbbm{Z}_2$ symmetry
allegedly responsible for dark-matter stability
cannot by itself alone
lead to any predictions for the lepton masses and mixings.
In section~3 I investigate some conditions
that the lepton flavour symmetry group $\mathcal{G}$
of a predictive model of the type of those in references~\cite{valle,morisi}
is likely to satisfy;
I show that $A_4$ is the smallest $\mathcal{G}$
satisfying those conditions,
but that there are other,
slightly larger groups which also satisfy them.
In section~4 I explicitly construct two further models
with an exact $\mathbbm{Z}_2$ symmetry
which is the remnant from a partially broken
lepton flavour symmetry group $\mathcal{G} \supset D_4$,
instead of $\mathcal{G} \equiv A_4$;
those models lead to realistic predictions for lepton mixing,
distinct from the ones of the models of references~\cite{valle,morisi}.
In section~5 I show that
the predictions of models of this type may be altered
when one considers the radiative contributions
to the neutrino mass matrix.
I summarize the main conclusions of this research in section~6.

\section{Non-existence of $\mathbbm{Z}_2$ symmetries
in the lepton sector}

We remind that
\be
U^T \mnu\, U = {\rm diag} \left( \mu_1, \mu_2, \mu_3 \right),
\label{diagmnu}
\ee
where the $\mu_j = m_j \exp{\left( 2 i \beta_j \right)}$
($j = 1, 2, 3$)
are the light-neutrino masses together with their respective Majorana phases.
The matrix $U$ is unitary.
The Majorana mass matrix $\mnu$ always has~\cite{grimusludl}
a $\mathbbm{Z}_2 \times \mathbbm{Z}_2 \times \mathbbm{Z}_2$ symmetry,
corresponding to the three columns $c_j$
of $U = \left( c_1, c_2, c_3 \right)$.
Indeed,
from equation~(\ref{diagmnu}),
\be
\mnu c_j = \mu_j c_j^\ast
\label{eigen}
\ee
for $j = 1, 2, 3$
(I do not use the summation convention in this section).
Unitarity of $U$ means that
\be
c_j^\dagger c_k = \delta_{jk}.
\label{unit}
\ee
Therefore,
from equation~(\ref{eigen}),
\be
\mnu = \sum_{j=1}^3 \mu_j c_j^\ast c_j^\dagger,
\label{mnew}
\ee
which is symmetric as it should.
We next define
\be
S \left( a_1, a_2, a_3 \right) = \sum_{j=1}^3 a_j c_j c_j^\dagger
\ee
with $a_{1,2,3} = \pm 1$.
The square of $S \left( a_1, a_2, a_3 \right)$ is the unit matrix
because of equation~(\ref{unit}) and of $a_1^2 = a_2^2 = a_3^2 = 1$.
Moreover,
\be
S \left( a_1, a_2, a_3 \right)^T \mnu\, S \left( a_1, a_2, a_3 \right) = \mnu,
\label{invariance}
\ee
with $\mnu$ given by equation~(\ref{mnew}).
Equation~(\ref{invariance}) means the invariance of $\mnu$ under
a $\mathbbm{Z}_2 \times \mathbbm{Z}_2 \times \mathbbm{Z}_2$
symmetry.
This invariance is completely independent
of the specific form of $\mnu$---though of course
the form of $S \left( a_1, a_2, a_3 \right)$
depends on the form of $\mnu$---and is just a mathematical consequence
of the mathematical fact that a symmetric matrix may be bi-diagonalized
by a unitary matrix.

However,
these three $\mathbbm{Z}_2$ invariances of $\mnu$ are \emph{not},
in general,
invariances of the (diagonal)
charged-lepton mass matrix $M_\ell$.\footnote{The exception would occur
if any of the three $c_j$ were either $\left( 1, 0, 0 \right)^T$,
$\left( 0, 1, 0 \right)^T$,
or $\left( 0, 0, 1 \right)^T$.
However,
this is not compatible with the experimental data
on the lepton mixing matrix $U$.}
This is because the charged-lepton masses are not degenerate.
For instance,
the mass matrix $\mnu$ may display $\mu$--$\tau$ reflection symmetry but,
since $m_\mu \neq m_\tau$,
that symmetry does not hold in the whole lepton sector.
Similarly~\cite{joshipura},
$U_{e3} = 0$ is equivalent to the presence
of a $\mathbbm{Z}_2$ symmetry in $\mnu$;
but,
once again,
that $\mathbbm{Z}_2$ cannot be extended to the charged-lepton sector.
\emph{There is no exact $\mathbbm{Z}_2$ symmetry
of the whole lepton sector}.
This means that the exact $\mathbbm{Z}_2$ symmetry
supposedly responsible for dark-matter stability cannot,
by itself alone,
explain any features of $\mnu$.

\section{An {\it Ansatz}\/ for predictive models}

If one contemplates the models
in references~\cite{valle,morisi},
one notices that in both of them
the three left-handed-lepton gauge-SU(2) doublets $D_{eL}$,
$D_{\mu L}$,
and $D_{\tau L}$ are \emph{invariant}
under the unbroken $\mathbbm{Z}_2$ subgroup
of the lepton flavour symmetry group $A_4$---they are not all
invariant under the full $A_4$
but are invariant under its $\mathbbm{Z}_2$ subgroup.

Let us remind that the group $A_4$ is discrete and non-abelian;
it has twelve elements,
one triplet irreducible representation (``irrep'')
and three inequivalent singlet representations.
In a convenient basis,
the triplet irrep of $A_4$ is generated by
\be
R = \left( \begin{array}{ccc}
1 & 0 & 0 \\ 0 & -1 & 0 \\ 0 & 0 & -1
\end{array} \right)
\quad {\rm and} \quad
T = \left( \begin{array}{ccc}
0 & 1 & 0 \\ 0 & 0 & 1 \\ 1 & 0 & 0
\end{array} \right).
\ee
$R$ generates the $\mathbbm{Z}_2$ subgroup of $A_4$ which stays unbroken.
In the models of references~\cite{valle,morisi},
the three left-handed-lepton doublets are each assigned
to one of the three inequivalent singlets of $A_4$,
\be
{\bf 1}_p: \quad R \to 1, \quad T \to e^{i\, \frac{2 \pi}{3}\, p}
\quad (p = 0, 1, 2),
\label{singlets}
\ee
which,
as stands evident in equation~(\ref{singlets}),
are invariant under the $\mathbbm{Z}_2$ subgroup of $A_4$ generated by $R$.

The predictions for $\mnu$ claimed in the models
of references~\cite{valle,morisi} do not result
from the unbroken $\mathbbm{Z}_2$ symmetry generated by $R$
and allegedly responsible for dark-matter stability.
Rather,
they follow from other features of the models,
{\it viz.}\ their restricted content of Higgs doublets
and right-handed neutrinos,
and the specific assignment
of those fields to representations of $A_4$.
It is those features,
neither the group $A_4$ nor its conserved subgroup $\mathbbm{Z}_2$
generated by $R$,
which make the models of references~\cite{valle} and~\cite{morisi}
differ from each other.

The question that then arises is whether it is possible
to construct other \emph{predictive} models,
beyond the ones in references~\cite{valle,morisi},
in which there is an unbroken $\mathbbm{Z}_2$
which is a subgroup of a larger,
but spontaneously broken,
lepton flavour symmetry group $\mathcal{G}$.
In particular,
we would like to ascertain whether $A_4$
is the only possible $\mathcal{G}$ of models of this type.

It is clear that any such model,
if it is to lead to any constraints on $\mnu$,
must treat the three left-handed neutrinos $\nu_{eL}$,
$\nu_{\mu L}$,
and $\nu_{\tau L}$ differently,
since the neutrino mass term is
\be
\frac{1}{2} \left( \nu_{eL}^T,\ \nu_{\mu L}^T,\ \nu_{\tau L}^T \right)
C^{-1} \mnu \left( \begin{array}{c} \nu_{eL} \\ \nu_{\mu L} \\ \nu_{\tau L}
\end{array} \right) + {\rm H.c.}
\ee
Now,
$\nu_{\alpha L}$ is a component of the SU(2) doublet $D_{\alpha L}$,
for $\alpha = e, \mu, \tau$,
hence the three $D_{\alpha L}$ must be treated differently in the model.
My {\it Ansatz}\/ is,
therefore,
that $D_{eL}$,
$D_{\mu L}$,
and $D_{\tau L}$ should sit in inequivalent \emph{singlet} representations
of $\mathcal{G}$.
On the other hand,
from the fact that
the unbroken $\mathbbm{Z}_2$ subgroup of $\mathcal{G}$ cannot constrain $\mnu$,
it follows that $D_{eL}$,
$D_{\mu L}$,
and $D_{\tau L}$ must be {\em invariant} under that $\mathbbm{Z}_2$.

We thus conclude that $\mathcal{G}$ should be a group
satisfying the following conditions:
\begin{enumerate}
\item It has a $\mathbbm{Z}_2$ subgroup.
\item It has some non-singlet representations
whose components transform differently---some components are even
and some others are odd---under the $\mathbbm{Z}_2$ subgroup.
This is necessary in order that $\mathcal{G}$ may be spontaneously broken
to the $\mathbbm{Z}_2$ subgroup when the $\mathbbm{Z}_2$-even
components of those representations acquire a
vacuum expectation value (VEV).
\item It has (at least) three inequivalent singlet representations.
\item Those three inequivalent singlet representations
are invariant under the $\mathbbm{Z}_2$ subgroup,
{\it i.e.}\ they represent trivially the $\mathbbm{Z}_2$ subgroup.
\end{enumerate}
A search \cite{ludl} for discrete groups satisfying these four constraints
shows that $A_4$ is the smallest of them.
There are,
though,
many other,
slightly larger groups satisfying the four constraints
and unrelated to $A_4$---for instance,
the group $D_4 \times \mathbbm{Z}_2$,
which has 16 elements and eight inequivalent singlets,
or the group $S_3 \times \mathbbm{Z}_3$,
which has 18 elements and six inequivalent singlets.

In the following section I shall present two models of my own,
based on lepton flavour symmetry groups $\mathcal{G}$
of the form $D_4 \times \mathbbm{Z}_n$.

\section{Other models}

\subsection{The group $D_n$}

The discrete non-abelian group $D_n$
($n \ge 2$)~\cite{lindner}
is generated by two transformations $A$ and $D$ which satisfy
\be
A^2 = D^2 = (A D)^n = e,
\label{Dndefinition}
\ee
where $e$ is the identity transformation.
The group $D_n$ has $2 n$ elements,
which are
\be
e,\ A,\ D,\ A D,\ D A,\ A D A,\ D A D,\ \ldots,\
(A D)^{n/2-1} A,\ (D A)^{n/2-1} D,\ (A D)^{n/2}
\ee
if $n$ is even,
and
\be
e,\ A,\ D,\ A D,\ D A,\ \ldots,\
(A D)^{(n-1)/2},\ (D A)^{(n-1)/2},\ (A D)^{(n-1)/2} A
\ee
if $n$ is odd.
The group $D_2$ is isomorphic to $\mathbbm{Z}_2 \times \mathbbm{Z}_2$.
The group $D_3$ is isomorphic to the permutation group of three elements,
$S_3$.

It follows from equation~(\ref{Dndefinition}) that
all the $D_n$ groups have two inequivalent singlet representations,
\ba
\mathbf{1}_{++}: & & A \to +1,\ D \to +1,
\\
\mathbf{1}_{--}: & & A \to -1,\ D \to -1.
\ea
($\mathbf{1}_{++}$ is the trivial representation.)
If $n$ is even,
then there are two further inequivalent singlet representations,
\ba
\mathbf{1}_{+-}: & & A \to +1,\ D \to -1,
\\
\mathbf{1}_{-+}: & & A \to -1,\ D \to +1.
\ea
All other irreps of $D_n$
(for $n \ge 3$)
are doublets.
They may be written,
in a real basis,
\be
{\bf 2}_q: \
A \to \left( \begin{array}{cc} 1 & 0 \\ 0 & -1 \end{array} \right),
\
D \to \left( \begin{array}{cc}
c_{nq} & s_{nq} \\ s_{nq} & - c_{nq} \end{array} \right),
\label{Dndoublets}
\ee
where $c_{nq} = \cos{\left( 2 q \pi / n \right)}$
and $s_{nq} = \sin{\left( 2 q \pi / n \right)}$,
for $q = 1, 2, \ldots, (n-2)/2$ if $n$ is even
and $q = 1, 2, \ldots, (n-1)/2$ if $n$ is odd.
In the cases of $D_3$ and $D_4$ one may drop the index $q$
because its unique value is then 1.

In $D_4$~\cite{we},
one has
\be
{\bf 2}: \
A \to \left( \begin{array}{cc} 1 & 0 \\ 0 & -1 \end{array} \right),
\
D \to \left( \begin{array}{cc}
0 & 1 \\ 1 & 0 \end{array} \right),
\ee
and
\be
\stackrel{{\bf 2}}{(a,\ b)}
\otimes
\stackrel{{\bf 2}}{(a',\ b')}
\ = \
\stackrel{{\bf 1}_{++}}{(a a' + b b')}
\oplus
\stackrel{{\bf 1}_{--}}{(a b' - b a')}
\oplus
\stackrel{{\bf 1}_{+-}}{(a a' - b b')}
\oplus
\stackrel{{\bf 1}_{-+}}{(a b' + b a')}.
\ee

In the representation of equation~(\ref{Dndoublets}),
$D_n$ is spontaneously broken,
but its $\mathbbm{Z}_2$ subgroup generated by $A$ stays unbroken,
when the upper component of a $D_n$ doublet acquires a VEV
but the lower component has null VEV;
it is furthermore necessary that
no singlet ${\bf 1}_{-\pm}$ acquires a VEV,
lest $A$ is broken by that VEV.
These will be my assumptions in the following models.

\subsection{A model using $D_4 \times \mathbbm{Z}_3$}
\label{model1}

The group $D_4 \times \mathbbm{Z}_3$ has 24 elements,
twice as many as $A_4$.
Still,
the following model has the advantage that it is economic in terms of fields,
since it uses only three right-handed neutrinos $\nu_{1,2,3R}$
and four ``Higgs'' gauge-SU(2) scalar doublets $\Phi_{1,2,3,4}$.
The fields transform under the family-symmetry group
according to table~\ref{tableD4Z3},
in which $\omega = \exp{\left( 2 i \pi / 3 \right)}$.
\begin{table}[h]
\begin{center}
\begin{tabular}{|c|c|c|}
\hline
$D_{eL}$ & $D_{\mu L}$ & $D_{\tau L}$ \\
$\left( {\bf 1}_{++},\ \omega^2 \right)$ &
$\left( {\bf 1}_{++},\ 1 \right)$ &
$\left( {\bf 1}_{+-},\ 1 \right)$ \\
\hline
$e_R$ & $\mu_R$ & $\tau_R$ \\
$\left( {\bf 1}_{++},\ \omega \right)$ &
$\left( {\bf 1}_{++},\ 1 \right)$ &
$\left( {\bf 1}_{+-},\ \omega^2 \right)$ \\
\hline
$\left( \nu_{1R},\ \nu_{2R} \right)$ & $\nu_{3R}$ & \\
$\left( {\bf 2},\ 1 \right)$ &
$\left( {\bf 1}_{++},\ 1 \right)$ &
 \\
\hline
$\left( \Phi_1,\ \Phi_2 \right)$ & $\Phi_3$ & $\Phi_4$ \\ 
$\left( {\bf 2},\ 1 \right)$ &
$\left( {\bf 1}_{++},\ 1 \right)$ &
$\left( {\bf 1}_{++},\ \omega \right)$ \\
\hline
\end{tabular}
\end{center}
\caption{Transformation rules of the fields
under the symmetry group $D_4 \times \mathbbm{Z}_3$.}
\label{tableD4Z3}
\end{table}
Notice that $D_{e,\mu,\tau L}$ are all $\mathbf{1}_{+\pm}$ under $D_4$,
therefore they are invariant under the $\mathbbm{Z}_2$ subgroup of $D_4$
generated by $A$
(the group $D_4 \times \mathbbm{Z}_3$ has 12 inequivalent singlets,
out of which six are invariant under that subgroup),
as prescribed by my {\it Ansatz}\/ of section~3.

The Yukawa Lagrangian is
\ba
\LL_{\rm Yuk} &=&
- \frac{m_e}{v_4}\, \bar D_{eL} e_R \Phi_4
- \frac{m_\mu}{v_3}\, \bar D_{\mu L} \mu_R \Phi_3
- \frac{m_\tau}{v_4}\, \bar D_{\tau L} \tau_R \Phi_4
\no & &
- \frac{a}{v_4^\ast}\, \bar D_{eL} \tilde \Phi_4 \nu_{3R}
- \frac{b}{v_3^\ast}\, \bar D_{\mu L} \tilde \Phi_3 \nu_{3R}
\no & &
- \frac{c}{v_1^\ast}\, \bar D_{\mu L} \left(
\tilde \Phi_1 \nu_{1R} + \tilde \Phi_2 \nu_{2R}
\right)
- \frac{d}{v_1^\ast}\, \bar D_{\tau L} \left(
\tilde \Phi_1 \nu_{1R} - \tilde \Phi_2 \nu_{2R}
\right)
+ {\rm H.c.},
\ea
where $\tilde \Phi_n = i \tau_2 \Phi_n^\ast$ for $n = 1, 2, 3, 4$
and $v_n$ is the VEV of $\Phi_n^0$
(the neutral component of $\Phi_n$).
One sees that the charged-lepton mass matrix is automatically diagonal
as a consequence of the flavour symmetry.
The neutrino Dirac mass matrix is
\be
M_D = \left( \begin{array}{ccc}
0 & 0 & a \\
c & \rho c & b \\
d & - \rho d & 0
\end{array} \right),
\label{MD2}
\ee
where $\rho = v_2^\ast / v_1^\ast$.
The Lagrangian of right-handed-neutrino Majorana masses is
\be
\LL_{\rm Maj} =
\frac{m_0}{2} \left( \nu_{1R}^T C^{-1} \nu_{1R}
+ \nu_{2R}^T C^{-1} \nu_{2R} \right)
+ \frac{m_1}{2}\, \nu_{3R}^T C^{-1} \nu_{3R}
+ {\rm H.c.}
\label{Maj}
\ee
Hence,
the right-handed-neutrino Majorana mass matrix is
\be
M_R = \left( \begin{array}{ccc}
m_0 & 0 & 0 \\
0 & m_0 & 0 \\
0 & 0 & m_1
\end{array} \right).
\label{MR2}
\ee
I assume that the $\mathbbm{Z}_2$ subgroup of $D_4$
generated by $A$ is left unbroken by the vacuum,
which means that
\be
v_2 = 0 \Leftrightarrow \rho = 0.
\label{preserve}
\ee
($\Phi_3$ and $\Phi_4$ are both $D_4$-invariant,
therefore their VEVs do not break $D_4$.)
Using the seesaw formula
\be
\mnu = - M_D M_R^{-1} M_D^T,
\label{seesaw}
\ee
it is easy to check that one obtains
\be
\left( \mnu \right)_{e\tau} = 0, \quad \det{\mnu} = 0.
\label{constraints2}
\ee
These are the predictions of the model.
It is not difficult to check that these predictions
are compatible with the experimental data
on neutrino masses and on lepton mixing.

\subsection{A model using $D_4 \times \mathbbm{Z}_4$}
\label{model2}

The following model also uses three right-handed neutrinos
and four Higgs doublets.
The fields transform under the family-symmetry group
according to table~\ref{tableD4Z4}.
\begin{table}[ht]
\begin{center}
\begin{tabular}{|c|c|c|}
\hline
$D_{eL}$ & $D_{2L}$ & $D_{3L}$ \\
$\left( {\bf 1}_{+-},\ 1 \right)$ &
$\left( {\bf 1}_{++},\ 1 \right)$ &
$\left( {\bf 1}_{++},\ -i \right)$ \\
\hline
$e_R$ & $\ell_{2R}$ & $\ell_{3R}$ \\
$\left( {\bf 1}_{+-},\ 1 \right)$ &
$\left( {\bf 1}_{++},\ 1 \right)$ &
$\left( {\bf 1}_{++},\ -i \right)$ \\
\hline
$\left( \nu_{1R},\ \nu_{2R} \right)$ & $\nu_{3R}$ & \\
$\left( {\bf 2},\ 1 \right)$ &
$\left( {\bf 1}_{++},\ -1 \right)$ &
 \\
\hline
$\left( \Phi_1,\ \Phi_2 \right)$ & $\Phi_3$ & $\Phi_4$ \\ 
$\left( {\bf 2},\ 1 \right)$ &
$\left( {\bf 1}_{++},\ 1 \right)$ &
$\left( {\bf 1}_{++},\ -i \right)$ \\
\hline
\end{tabular}
\end{center}
\caption{Transformation rules of the fields
under the symmetry group $D_4 \times \mathbbm{Z}_4$.}
\label{tableD4Z4}
\end{table}
In this model the charged-lepton mass matrix is not automatically diagonal
as a consequence of the symmetry;
the right-handed charged leptons $\mu_R$ and $\tau_R$
are linear combinations of $\ell_{2R}$ and $\ell_{3R}$,
and the left-handed-lepton doublets $D_{\mu L}$ and $D_{\tau L}$
are linear combinations of $D_{2L}$ and $D_{3L}$.
The Yukawa Lagrangian is
\ba
\LL_{\rm Yuk} &=&
- \frac{m_e}{v_3}\, \bar D_{eL} e_R \Phi_3
- \frac{r}{v_3}\, \bar D_{2L} \ell_{2R} \Phi_3
- \frac{s}{v_4}\, \bar D_{3L} \ell_{2R} \Phi_4
- \frac{t}{v_3}\, \bar D_{3L} \ell_{3R} \Phi_3
\no & &
- \frac{f}{v_1^\ast}\, \bar D_{eL} \left(
\tilde \Phi_1 \nu_{1R} - \tilde \Phi_2 \nu_{2R}
\right)
- \frac{g}{v_1^\ast}\, \bar D_{2L} \left(
\tilde \Phi_1 \nu_{1R} + \tilde \Phi_2 \nu_{2R}
\right)
\no & &
- \frac{h}{v_4^\ast}\, \bar D_{3L} \tilde \Phi_4 \nu_{3R}
+ {\rm H.c.}
\ea
The charged-lepton mass matrix is
\be
M_\ell = \left( \begin{array}{ccc}
m_e & 0 & 0 \\
0 & r & 0 \\
0 & s & t
\end{array} \right).
\ee
The neutrino Dirac mass matrix is
\be
M_D = \left( \begin{array}{ccc}
f & - \rho f & 0 \\
g & \rho g & 0 \\
0 & 0 & h
\end{array} \right).
\label{MD3}
\ee
Equations~(\ref{Maj}) and~(\ref{MR2}) also hold in this model,
and we suppose as before that the vacuum features equation~(\ref{preserve}).
One then obtains the predictions
\be
m_1 = 0, \quad U_{e3} = 0.
\label{constraints3}
\ee
The second of these predictions is at present
somewhat disfavoured~\cite{schwetz}
but the situation is not settled yet.

\section{Radiative neutrino masses}

\subsection{The scalar potentials}

The scalar potential for the models of section~4 is
\be
V_{D_4} = V_{\rm sym} + V_{\rm sb},
\ee
where
\be
V_{\rm sym} =
\cdots
+ \lambda_{12} \left[ \left( \Phi_1^\dagger \Phi_2 \right)^2
+ {\rm H.c.} \right]
+ \left\{ \lambda_{13} \left[ \left( \Phi_1^\dagger \Phi_3 \right)^2
+ \left( \Phi_2^\dagger \Phi_3 \right)^2 \right]
+ {\rm H.c.} \right\}
\label{VD4}
\ee
is invariant under a $D_4 \times \mathbbm{Z}_3$
or $D_4 \times \mathbbm{Z}_4$ transformation,
and
\be
V_{\rm sb} = \mu_4 \Phi_3^\dagger \Phi_4 + {\rm H.c.}
\ee
breaks softly the $\mathbbm{Z}_{3,4}$ while preserving $D_4$.
The term in $V_{\rm sb}$ is needed lest a neutral component of $\Phi_4$
becomes a Goldstone boson,
since the discrete $\mathbbm{Z}_3$ symmetry of the first model,
or $\mathbbm{Z}_4$ of the second model,
effectively becomes,
in $V_{\rm sym}$,
a continuous U(1) symmetry.
The parameters $\lambda_{13}$ and $\mu_4$ are complex,
$\lambda_{12}$ is real;
however,
the phases of $\Phi_3$ and $\Phi_4$ may be rotated
in such a way that $\lambda_{13}$ and $\mu_4$ become real too.

The models of references~\cite{valle,morisi}
feature an $A_4$ triplet $\left( \Phi_1, \Phi_2, \Phi_3 \right)^T$
of Higgs doublets,
together with a fourth Higgs doublet,
$\Phi_4$,
which is $A_4$-invariant.
Their scalar potential is
\ba
V_{A_4} &=& \cdots
+ \left\{ \lambda_7 \left[
\left( \Phi_1^\dagger \Phi_2 \right)^2
+ \left( \Phi_2^\dagger \Phi_3 \right)^2
+ \left( \Phi_3^\dagger \Phi_1 \right)^2
\right]
\right. \no & &
+ \lambda_8 \left[
\left( \Phi_1^\dagger \Phi_4 \right)^2
+ \left( \Phi_2^\dagger \Phi_4 \right)^2
+ \left( \Phi_3^\dagger \Phi_4 \right)^2
\right]
\no & &
+ \lambda_9 \left(
  \Phi_2^\dagger \Phi_3\, \Phi_1^\dagger
+ \Phi_3^\dagger \Phi_1\, \Phi_2^\dagger
+ \Phi_1^\dagger \Phi_2\, \Phi_3^\dagger
\right) \Phi_4
\no & & \left.
+ \lambda_{10} \left(
  \Phi_3^\dagger \Phi_2\, \Phi_1^\dagger
+ \Phi_1^\dagger \Phi_3\, \Phi_2^\dagger
+ \Phi_2^\dagger \Phi_1\, \Phi_3^\dagger
\right) \Phi_4
+ {\rm H.c.} \right\},
\label{VA4}
\ea
where $\lambda_{7,8,9,10}$ are complex.
In this case,
no soft-breaking term is necessary.

In equations~(\ref{VD4}) and~(\ref{VA4}),
the ``$\cdots$'' stand for terms of the forms
\be
\Phi_n^\dagger \Phi_n, \quad
\left( \Phi_n^\dagger \Phi_n \right)^2, \quad
\left( \Phi_n^\dagger \Phi_n \right)
\left( \Phi_{n^\prime}^\dagger \Phi_{n^\prime} \right), \quad
\left( \Phi_n^\dagger \Phi_{n^\prime} \right)
\left( \Phi_{n^\prime}^\dagger \Phi_n \right)
\ee
which are not relevant for the purposes of this section.

\subsection{The one-loop diagram in the $D_4$ models}

The potentials of the previous subsection cause
a radiative generation of neutrino masses,
via a one-loop diagram~\cite{ma},
which partially destroys the predictive power of the models.
In equation~(\ref{VD4}),
the crucial terms are $\left( \Phi_2^\dagger \Phi_1 \right)^2$,
$\left( \Phi_2^\dagger \Phi_3 \right)^2$,
and their Hermitian conjugates.
Indeed,
even when the VEV of $\Phi_2^0$ vanishes,
as assumed here,
those terms allow one to substitute the seesaw mass term,
obtained via the insertion of two VEVs of $\Phi_2^0$,
by a radiative mass term in which a $\Phi_2^0$ runs in the loop,
connected to two $\Phi_1^0$ or $\Phi_3^0$,
respectively,
and to an insertion of their VEVs.

The one-loop radiative corrections to the neutrino mass matrix in seesaw models
have been studied in reference~\cite{radiative}.
The final result of that paper is that,
to the neutrino mass matrix in equation~(\ref{seesaw}),
a radiative contribution must be added which reads---taking into account
that the mass matrix of the heavy neutrinos,
$M_R$ in equation~(\ref{MR2}),
is diagonal---
\be
\frac{3 g^2}{32 \pi^2 c_w^2}\,
M_D \frac{\ln{\left( M_R / m_Z \right)}}{M_R} M_D^T
+ \sum_b \frac{m_b^2}{16 \pi^2}\,
\sum_{n, n^\prime} b_n b_{n^\prime}
\Delta_n \frac{\ln{\left( M_R / m_b \right)}}{M_R} \Delta_{n^\prime}^T.
\label{final}
\ee
The first term of this equation
is the contribution from the diagram with a gauge boson $Z^0$ in the loop;
this is very much similar to the tree-level seesaw equation~(\ref{seesaw})
and needs not concern us.
In the second term of equation~(\ref{final}),
the index $b$ denotes the seven physical
(mass eigenstate)
neutral scalar bosons of the model,
$S_b^0$,
which have masses $m_b$~\cite{scalars}.
The complex quantities $b_n$ satisfy
\be
\Phi_n^0 = v_n + \frac{1}{\sqrt{2}}\, \sum_b b_n S_b^0,
\ee
for $n = 1, 2, 3, 4$.
The $3 \times 3$ matrices $\Delta_n$
are the matrices of the Yukawa couplings of each $\Phi_n$,
{\it viz.},
in the model of subsection~4.2,
\be
\begin{array}{rclcrcl}
\Delta_1 &=& {\displaystyle \frac{1}{v_1^\ast}} \left( \begin{array}{ccc}
0 & 0 & 0 \\ c & 0 & 0 \\ d & 0 & 0
\end{array} \right),
& &
\Delta_2 &=& {\displaystyle \frac{1}{v_1^\ast}} \left( \begin{array}{ccc}
0 & 0 & 0 \\ 0 & c & 0 \\ 0 & -d & 0
\end{array} \right),
\\
\Delta_3 &=& {\displaystyle \frac{1}{v_3^\ast}} \left( \begin{array}{ccc}
0 & 0 & 0 \\ 0 & 0 & b \\ 0 & 0 & 0
\end{array} \right),
& &
\Delta_4 &=& {\displaystyle \frac{1}{v_4^\ast}} \left( \begin{array}{ccc}
0 & 0 & a \\ 0 & 0 & 0 \\ 0 & 0 & 0
\end{array} \right).
\end{array}
\ee
It is easy to see that the term
\be
\sum_b \frac{m_b^2}{16 \pi^2}\,
b_2^2\,
\Delta_2\, \frac{\ln{\left( M_R / m_b \right)}}{M_R}\, \Delta_2^T
= \sum_b \frac{m_b^2 b_2^2}{16 \pi^2 {v_1^\ast}^2}\,
\frac{\ln{\left( m_0 / m_b \right)}}{m_0}
\left( \begin{array}{ccc}
0 & 0 & 0 \\ 0 & c^2 & - c d \\ 0 & - c d & d^2
\end{array} \right)
\label{final2}
\ee
is an extra contribution to the $\mnu$ of equation~(\ref{seesaw})
that destroys the prediction $\det{\mnu} = 0$
in equation~(\ref{constraints2}).
On the other hand,
the prediction $\left( \mnu \right)_{e \tau} = 0$ in the same equation
remains unaffected
(at least at the one-loop level).

The practical effect of the terms $\left( \Phi_2^\dagger \Phi_1 \right)^2$,
$\left( \Phi_2^\dagger \Phi_3 \right)^2$,
and their Hermitian conjugates,
in equation~(\ref{VD4}),
is to split the degeneracy of the real and imaginary components of $\Phi_2^0$.
If one writes $\Phi_2^0 = \rho_2 + i \eta_2$,
then the relevant mass terms are $A_{22} \rho_2^2 + B_{22} \eta_2^2$
with $A_{22} \neq B_{22}$.
The quantity $\sum_b m_b^2 b_2^2 = A_{22} - B_{22}$ in equation~(\ref{final2})
is non-vanishing because of those terms in the scalar potential.

Similarly,
in the model of subsection~4.3,
\be
\begin{array}{rclcrcl}
\Delta_1 &=& {\displaystyle \frac{1}{v_1^\ast}} \left( \begin{array}{ccc}
f & 0 & 0 \\ g & 0 & 0 \\ 0 & 0 & 0
\end{array} \right),
& &
\Delta_2 &=& {\displaystyle \frac{1}{v_1^\ast}} \left( \begin{array}{ccc}
0 & -f & 0 \\ 0 & g & 0 \\ 0 & 0 & 0
\end{array} \right),
\\
\Delta_3 &=& \left( \begin{array}{ccc}
0 & 0 & 0 \\ 0 & 0 & 0 \\ 0 & 0 & 0
\end{array} \right),
& &
\Delta_4 &=& {\displaystyle \frac{1}{v_4^\ast}} \left( \begin{array}{ccc}
0 & 0 & 0 \\ 0 & 0 & 0 \\ 0 & 0 & h
\end{array} \right).
\end{array}
\ee
The relevant term is
\be
\sum_b \frac{m_b^2}{16 \pi^2}\,
b_2^2\,
\Delta_2\, \frac{\ln{\left( M_R / m_b \right)}}{M_R}\, \Delta_2^T
= \sum_b \frac{m_b^2 b_2^2}{16 \pi^2 {v_1^\ast}^2}\,
\frac{\ln{\left( m_0 / m_b \right)}}{m_0}
\left( \begin{array}{ccc}
f^2 & - f g & 0 \\ - f g & g^2 & 0 \\ 0 & 0 & 0
\end{array} \right)
\ee
and destroys the prediction $m_1 = 0$ in equation~(\ref{constraints3}).
The other prediction in that equation,
{\it viz.}\ $U_{e3} = 0$,
remains unaffected,
though.

The fact that the one-loop radiative contributions
to the neutrino mass matrix
are suppressed by loop factors $\left( 16 \pi^2 \right)^{-1}$
renders,
in general,
their practical consequences small.
Take for instance the model of subsection~4.3:
the masses of the first two light neutrinos,
$m_1$ and $m_2$,
will be generated by the $2 \times 2$ mass matrix
\be
\frac{1}{m_0}
\left( \begin{array}{cc}
f^2 \left( 1 + \epsilon \right) & f g \left( 1 - \epsilon \right) \\
f g \left( 1 - \epsilon \right) & g^2 \left( 1 + \epsilon \right)
\end{array} \right),
\ee
where $\epsilon \propto \left( 16 \pi^2 \right)^{-1}$ is small.
In the limit of small $\epsilon$,
\be
m_1 \approx \frac{\left| \epsilon \right|}{m_0}\,
\frac{4 \left| f g \right|^2}{\left| f \right|^2 + \left| g \right|^2},
\quad
m_2 \approx \frac{\left| f \right|^2 + \left| g \right|^2}{m_0},
\ee
so that $m_1 / m_2 \le \left| \epsilon \right| \sim 10^{-2}$.

\subsection{The one-loop diagram in the $A_4$ models}

The situation with the $A_4$ models of references~\cite{valle,morisi}
is more complicated,
because in those models there are two,
instead of only one,
VEV-less Higgs doublets $\Phi_2$ and $\Phi_3$.
In the scalar potential there are terms
$\left( \Phi_{1,4}^\dagger \Phi_2 \right)^2$,
$\left( \Phi_{1,4}^\dagger \Phi_3 \right)^2$,
$\Phi_1^\dagger \Phi_2 \, \Phi_4^\dagger \Phi_3$,
and $\Phi_4^\dagger \Phi_2 \, \Phi_1^\dagger \Phi_3$,
which lead to one-loop diagrams that mimic the situation
in which both $\Phi_2$ and $\Phi_3$ actually had a VEV.
Furthermore,
those terms lead to CP violation
via mixing of the real and imaginary parts of $\Phi_2^0$ and $\Phi_3^0$.
All the predictions for the neutrino masses and lepton mixings
of those two models in general get lost,
but the corrections to those predictions are suppressed by loop factors
and should be small.

\section{Conclusions}

In this paper I have commented
on two recently proposed models~\cite{valle,morisi}
which make a connection between
some predictions for lepton mixing
and an unbroken $\mathbbm{Z}_2$ subgroup
of the lepton flavour symmetry group $A_4$.
I have found that:
\begin{itemize}
\item The predictions for lepton mixing claimed in those models
are not exclusively a consequence of the unbroken $\mathbbm{Z}_2$ subgroup.
Rather,
they follow mainly from the limited spectrum
of right-handed neutrinos and of Higgs doublets in those models,
and from the assignments of those fields to specific representations of $A_4$.
\item There are many other lepton flavour symmetry groups $\mathcal{G}$,
beyond $A_4$,
which may be used for constructing models
with features analogous to the ones of those commented upon.
The group $A_4$ is the smallest possible $\mathcal{G}$,
but any other group with a $\mathbbm{Z}_2$ subgroup
into which it may be spontaneously broken,
and with three inequivalent singlets
which transform trivially under that $\mathbbm{Z}_2$ subgroup,
can in principle be useful.
\item The predictions claimed in those two models are altered
when one considers the one-loop diagrams
which generate radiative contributions
to the neutrino mass matrix.
Those one-loop diagrams mimic non-zero VEVs
for the Higgs doublets that,
in those models,
are assumed wo have vanishing VEV.
\end{itemize}
I have also presented two further models of a similar type,
but which use symmetry groups $\mathcal{G} = D_4 \times \mathbbm{Z}_n$
instead of $\mathcal{G} = A_4$.
Both those models predict (after radiative corrections)
and almost massless neutrino.
Those models display other predictions
which are \emph{not} disturbed by the one-loop contribution
to the neutrino mass matrix.
In those models,
there are one Higgs doublet and one right-handed neutrino
which are odd under the conserved $\mathbbm{Z}_2$ symmetry
and may contribute to dark matter.

\vspace*{10mm}

\paragraph{Acknowledgements:} I gratefully thank Patrick Otto Ludl
for performing for me systematic searches for discrete symmetry groups
satisfying various criteria.
I also thank Ernest Ma for calling my attention
to his mechanism of radiative neutrino mass generation
and its consequences.
I thank the organizers,
notably Bohdan Grzadkowski and Maria Krawczik,
of the conference {\it Scalars 2011},
which took place in Warsaw under a very pleasant atmosphere,
for allowing me to present and discuss this work at that conference.
Last but not least,
I thank Walter Grimus for reading the manuscript
and making useful criticisms and suggestions.
My work is funded by the Portuguese
{\it Funda\c c\~ao para a Ci\^encia e a Tecnologia} (FCT)
through FCT unit 777 and the projects CERN/FP/116328/2010,
PTDC/FIS/098188/2008,
and PTDC/FIS/117951/2010,
and also by the Marie Curie Initial Training Network ``UNILHC''
PITN-GA-2009-237920.

\end{document}